\documentstyle[12pt]{article}

\def\be{\begin{equation}}
\def\ee{\end{equation}}
\def\bea{\begin{eqnarray}}
\def\eea{\end{eqnarray}}
\def\vk{\bf k}
\def\vko{{\bf k}_1}
\def\vkt{{\bf k}_2}

\begin{document}

\begin{titlepage}

\begin{flushright}
Preprint\\
FIAN/TD-22/98
\end{flushright}

\begin{centering}

\vfill

{\bf INFLUENCE OF THE SOURCE EVOLUTION \\ON PARTICLE CORRELATIONS}\\

\vspace{1cm}

 I.V.ANDREEV\footnote{Postal address:Theory Dept.,P.N.Lebedev Institute,
                      Leninsky pr.53, 117924, Moscow, Russia\\
                      E-mail address: andreev@lpi.ac.ru\\
                      Tel: 7 095 1326743, Fax: 7 095 1358553\\}
\vspace{0.5cm}

{\it
P.N.Lebedev Physical Institute, Moscow, Russia}\\
\vspace{3cm}
\centerline{\bf Abstract}
\end{centering}
\vspace{0.3cm}\noindent

Modification of the particles
in the course of the source evolution is considered. Influence of this
effect on multiplicities and correlations of the particles is displayed,
including an enhancement of the production rates and identical particle
 correlations
and also back-to-back particle-antiparticle correlations.

\vspace{2cm}\noindent
Classification codes: 13.85.Hd, 13.85.Ni

\vfill \vfill

\end{titlepage}

\section{Introduction}

In this paper we consider effects which arise when particles are initially
produced inside hadronic matter (particle source) and only afterwards,
at the very last stage of the process, fly away as free particles. Being
produced in the hadronic medium they represent a part of the medium,
rather quasiparticles than free particles. So their spectrum $E(\bf k)$
changes in the course of the source evolution (say expansion).

Such a picture leads to specific observable effects. First, production rates
and identical pion correlations (HBT effect) can be amplified significantly.
Second, back-to-back particle-antiparticle correlations (PAC) appear
which are not necessarily small if quasiparticles differ essentially
from free particles.

Let us remind in this connection that momentary transformation of
quasiparticles into free particles was first considered~\cite{AW} in the
frame of simple oscillator model. Chaotic squeezed states of the particles
arise in this model giving rise to enhancement of single-particle inclusive
production and identical particle correlations, but antiparticles
(and so PAC) are absent in the model. On the other hand, the presence
of back-to-back PAC was noted~\cite{APW} some time ago for particle
production in the vacuum. However in this case PAC are suppressed (no
source evolution effect in this case). Unsuppressed back-to-back correlations
arising due to transformation of neutral quasiparticles into free particles
were found recently in ref.~\cite{CS}. This kind of correlations (including
charged pions) was also suggested quite recently~\cite{HM} as a probe
of disoriented chiral condensate (DCC) formation. Here we consider pion
creation inside a finite size source which undergoes time evolution
(having finite lifetime). Both kinds of effects will be found, irrespective
of DCC formation.

Below we consider multiple production processes at high energies. It is
suggested that after collision an excited volume is formed which undergoes
evolution. The last (hadronic) stage of the evolution is under consideration.
So the specific effects discussed below are expected to be absent in those
scenarios of particle production which do not treat the hadronization
stage in detail. In particular one can hardly expect these effects in
existing event generators though they are natural in hydrodynamical model.

\section{ Neutral pion production and evolution}

Neutral pions are represented by real valued field $\varphi({\bf x},t)$.
Its decomposition has the form

\be
\varphi({\bf x},t)=\int \frac{d^{3}k}{(2\pi)^{3/2}(2E_{k})^{1/2}}
\left[ a_{\vk}e^{-iEt+i{\vk}{\bf x}}+a^{\dag}_{\vk}e^{iEt-i{\vk}{\bf x}}
\right]
\label{eq:1}
\ee
where annihilation and creation operators $a_{\vk},a^{\dag}_{\vk}$
satisfy canonical commutation relations.

To describe the field emission we introduce an effective classical current
 $j({\bf x},t)$ (it is the practice in HBT effect study~\cite{{APW},{GKW}}).
The currents encode the space-time region of the particle production
including time duration of the process. Clearly the currents represent some
other particles producing pions. The current is considered as a random function
and subject of averaging (chaotic source). Production cross-sections and
correlations of identical particles can be now expressed through the mass
shell Fourier transform of the currents:

\be
j(E_{k},{\vk})=\int dt e^{iE_{k}t}j_{\vk}(t)=
\int d^{4}x e^{iE_{k}t-i{\vk}{\bf x}}j({\bf x},t)
\label{eq:2}
\ee
In particular the field correlators are:

\be
< a^{\dag}_{\vko}a_{\vkt}>_{0}=
\frac{1}{2\sqrt{E_{1}E_{2}}}
< j^{*}(E_{1},{\vk}_{1})j(E_{2},{\vk}_{2})>
\label{eq:3}
\ee
and, up to small corrections,

\be
<a_{\vko}a_{\vkt}>_{0}=0
\label{eq:4}
\ee
where brackets include statistical averaging and index $0$ in {\em lhs} of
the equations indicates that the medium evolution is not taken into account.

If production takes place in a medium one has to modify the free propagation
of the fields, say introduce some external potential. In our problem
the time evolution of the medium is essential, so we confine ourselves
to the homogenious sources. Then the problem is reduced to propagation
of the quasiparticles having variable energy, $E_{k}=E_{k}(t)$. For
example one may introduce variable mass $m(t)$ changing in the course
of evolution with $m({\infty})=m_{\pi}$.

Then our model is given by the Hamiltonian

\be
H=\frac{1}{2}\left[\pi^{2}(x)+({\bf\bigtriangledown}\varphi)^{2}(x)
+m^{2}(t)\varphi^{2}(x)\right]-j(x)\varphi(x),\qquad   \pi=\dot \varphi
\label{eq:5}
\ee
This Hamiltonian is similar to that of quantum oscillator with variable
frequency in the presence of external force (which is the current now).
Solution of the last problem is known for a long time~\cite{H}.
It is helpful to represent the solution of the equations of motion
in the form of canonical transformation of the creation and annihilation
operators. The only new feature of the field theory model given by Eq.5
(in comparison with nonstationary quantum oscillator) is an appearance of two
modes ${\vk}$ and ${-\vk}$ involved in the canonical transformation
(as it was the case in the original Bogolubov transformation from particles
to quasiparticles in superfluid~\cite{B}).

The resulting Bogolubov transformation solving the model given by Eq.5
has the form:

\be
\left( a_{\bf k}(t)\atop a^{\dag}_{-\bf k}(t)\right) =
\left( u \qquad v \atop v^{*} \qquad u^{*} \right)
\left( a_{\bf k}(0)+d_{\bf k}(t) \atop a^{\dag}_{-\bf k}(0)
+ d^{*}_{-{\bf k}}(t) \right)
\label{eq:6}
\ee
with

\be
d_{\bf k}(t)=\frac{i}{{(2E_{k})}^{1/2}} \int\limits_0^t dt_1
                                             \xi(t_{1})j_{\bf k}(t_{1})
\label{eq:7}
\ee
where $\xi(t)$ is the classical solution for the oscillator with variable
frequency and with initial conditions

\bea
\xi(0)=1, \qquad  \dot \xi (0)=iE_{k}(0)
\label{eq:8}
\eea
(we suggest that the currents are absent at $t<0$)

Coefficients of the transformation satisfy an equation

\be
|u|^2 -|v|^2 =1
\label{eq:9}
\ee
and can be expressed through the same function $\xi(t)$.
So the form of $\xi(t)$ is quite essential. In general it is an oscillating
function with variable frequency and amplitude. What is more important,
time reflected wave appears in $\xi(t)$. In particular keeping $E(t)$
constant in a small interval near some intermediate point $t=\bar t$
one gets

\bea
{(2\bar E)}^{-1/2}\xi(\bar t)=\bar u^{*}e^{i\bar E\bar t}
+\bar{v}e^{-i\bar E\bar t},
\qquad \bar E=E(\bar t)
\label{eq:10}
\eea
where $\bar u,\bar v$ are Bogolubov coefficients taken at $t=\bar t$
(with their oscillating time dependence canceled).

The appearance of time-reflected wave leads, throught Eq.7
to two types of contributions to observable quantities like $<a^{\dag}a>$.
The first is of the form
\be
\int\int dt_{1}dt_{2}e^{i(E_{1}t_{1}-E_{2}t_{2})}j(t_{1})j(t_{2})
\label{eq:11}
\ee
and the second is
\be
\int\int dt_{1}dt_{2}e^{i(E_{1}t_{1}+E_{2}t_{2})}j(t_{1})j(t_{2})
\label{eq:12}
\ee
The contributions of the first type lead to unsuppressed correlations
which depend on the difference $E_{1}-E_{2}$ being maximal at $E_{1}=E_2$.
The second type contributions depend on the sum $E_{1}+E_{2}$
(see ref.~\cite{APW}) and they are always suppressed. Below we neglect
the second type contributions keeping only "large" ones (of the first
type). Note that expectation $<aa>$ also contains large contributions.

To get simple explicit results we simplify the above expressions.
Firstly, the Bogolubov coefficients are taken to be real valued with
\be
u=\cosh r, \qquad v=\sinh r
\label{eq:13}
\ee
Secondly, the solution $\xi(t)$ is taken at a middle point $\bar t$,
see Eq.10. We also suggest that the field expectations are absent at the
initial moment $t=0$ (no quasiparticles up to $t=0$; presumably this is
the moment of the phase transition to the hadronic phase).
 Then, using Eqs.6,7 we obtain
expectation values at $t=\infty$:
\bea
{\langle a^{\dag}({\vko})a({\vkt})\rangle}=
\cosh(r_{1}+r_{2}-\bar r_{1}-\bar r_{2})\frac{1}{2\sqrt{\bar{E_{1}}\bar{E_{2}}}}
\langle j^{*}(\bar E_{1},{\vko})j(\bar E_{2},{\vkt})\rangle \nonumber \\
+\sinh r_{1}\sinh r_{2} F({\vko}-{\vkt})
\label{eq:14}
\eea

\bea
{\langle a({\vko})a({\vkt})\rangle}_{j}=
\sinh(r_{1}+r_{2}-\bar r_{1}-\bar r_{2})\frac{1}{2\sqrt{\bar{E_{1}}\bar{E_{2}}}}
\langle j^{*}(\bar E_{1},{\vko})j(\bar E_{2},-{\vkt})\rangle \nonumber \\
+\frac{1}{2}\sinh(r_{1}+r_{2})F({\vko}+{\vkt})
 \label{eq:15}
\eea
to be compared with Eqs.3,4.

In Eqs.14,15 $r_i$ are full evolution parameters and $\bar r_{i}$ are the
parameters evaluated for medium evolution from initial moment $t=0$ up
to a middle point $t=\bar t$, see Eq.10 (roughly $\bar r =r/2$).
The first terms in {\em rhs} of Eqs.14,15 represent contributions of the
pions produced by the currents (that is by some other particles) which
are now enhanced in comparison with no evolution production, Eqs.3,4.
The second terms in {\em rhs} of these equations contain form-factor
$F(\vk)$ which is Fourier transform of the initial volume $V$,
\be
F(0)=V/(2\pi)^{3}
\label{eq:16}
\ee
These terms represent the result of decay of the ground state of the medium.
In another language, they represent parametric excitation of the field
oscillators existing side by side with forced excitation if the oscillator
parameters depend on time.

The evolution effect is determined by the evolution parameter $r$ (the
effect vanishes if $r=0$).The value of $r$ for every momentum $\vk $ depends
in general on initial energy $E_{in}$, final energy $E_{f}$ and characteristic
time duration $T$ of the hadronic stage of the process. The parameter $r$
is maximal for small time duration

\be
r_{max}=\frac{1}{2}\ln \left( \frac{E_{f}}{E_{in}} \right) ,\qquad
ET\ll 1
\label{eq:17}
\ee
and vanishes for $ET \gg 1$ (adiabatic process).

As the final result we obtain expressions for inclusive production of
neutral pions. Single-particle distribution is
\bea
\frac {1}{\sigma}\frac{d\sigma}{d^{3}k}({\vk})
=\langle a^{\dag}({\vk})a({\vk}) \rangle
=\cosh(2r-2\bar r) \frac {1}{2\bar E} \langle |j(\bar {E},{\vk})|^2 \rangle
+\sinh^{2}r\frac{V}{(2\pi)^3}
\label{eq:18}
\eea
where the first term is the usual production rate (up to pion energy
modification) enhanced by $\cosh $ factor arising due to source
evolution. The second term gives an additional contribution due to
ground state decay.

Two-particle inclusive distribution under consideration is given by
\bea
\frac{1}{\sigma}\frac{d^{2}\sigma}{d^{3}k_{1}d^{3}k_{2}}
=\langle a^{\dag}_{1}a^{\dag}_{2}a_{1}a_{2} \rangle
=\langle a^{\dag}_{1}a_{1}\rangle \langle a^{\dag}_{2}a_{2}\rangle
+\langle a^{\dag}_{1}a_{2}\rangle \langle a^{\dag}_{2}a_{1}\rangle
+\langle a^{\dag}_{1}a^{\dag}_{2}\rangle \langle a_{1}a_{2}\rangle
\label{eq:19}
\eea
where the expectations in $\em rhs$ are given by Eqs.14,15. The first
term in $\em rhs$ of Eq.19 is the product of single-particle
distributions, the second term gives HBT effect and the third
term describes back-to-back particle-antiparticle correlatios
(PAC,see below) arising due to source evolution.

The relative correlation function which is measured in experiment is given by
\be
C({\vko,\vkt})=\frac{\sigma d^{2}\sigma/d^{3}k_{1}d^{3}k_{2}}
                  {(d\sigma/d^{3}k_{1})(d\sigma/d^{3}k_{2})}=
               1+C_{HBT}({\vko,\vkt}) + C_{PAC}({\vko,\vkt})
\label{eq:20}
\ee

The function $C_{HBT}({\vko,\vkt})$, describing HBT effect, reaches its
maximum at ${\vko}-{\vkt}=0$ where
\be
C_{HBT}({\vk,\vk})=1
\label{eq:21}
\ee
as usual though the slope of its momentum difference dependence is modified
(it is diminished due to $r$-dependence, that is due to evolution effect,
as one can see from Eqs.14,18).

The function $C_{PAC}({\vko,\vkt})$ describes PAC effect and gives an
additional positive contribution to the correlation, this contribution
being maximal at ${\vko+\vkt}=0$. The width of the PAC peak is expected
to be close to that of HBT peak (compare Eq.14 and Eq.15). Its height
depends crucially on evolution parameter $r$, vanishing at $r=0$ (see Eq.14)
that is in the absence of the medium effect. The influence of the PAC effect
on the function $C({\vko,\vkt})$ in Eq.20 at small momentum differences
may be noticeable for soft neutral pions having small momenta ${\vko,\vkt}$
and consequently small momentum differences ${\vko-\vkt}$. So one may expect
an increase of the correlation function $C({\vko,\vkt})$ at all momenta
${\bf k}_{i}$ for soft neutral pions having $|{\vk}|\le1/R$ where $R$ is
a characteristic size of the source. In particular from Eqs.14,15 and
Eqs.18-20 we get an estimate:

\be
C(0,0)\approx2+\left|\frac{\left(\frac{1}{\sigma}\frac{d\sigma}{d^{3}k}\right)_{0}({\vk}=0)\sinh r
+\frac{1}{2}\frac{V}{(2\pi)^{3}}\sinh2r}
                          {\left(\frac{1}{\sigma}\frac{d\sigma}{d^{3}k}\right)_{0}({\vk}=0)\cosh r
        + \frac{V}{(2\pi)^{3}}\sinh^{2}r} \right|^{2}
\label{eq:22}
\ee
where we put $\bar r=r/2$ and neglected pion energy modification
in the current expectations so that the differential cross-sections in Eq.22
having index $0$ are those without the evolution effect. Analogous
equation is valid for ${\pi}^{+}-{\pi}^{-}$ correlations (see below).
Numerical estimations of the
PAC effect for different colliding particles will be given elsewhere.

\section{Charged pion production}

Consideration of charged pions is analogous to that of neutral ones.
Now the field is complex valued and its decomposition reads

\bea
\phi (x)=\int \frac{d^{3}k}{(2\pi)^{3/2}(2E_{k})^{1/2}}
 \left[ a_{\vk}e^{-iEt+i{\vk}{\bf x}}+b^{\dag}_{\vk}e^{iEt-i{\vk}{\bf x}}\right]
\label{eq:23}
\eea
containing annihilation operators $a_{\vk}$ for particles (say ${\pi}^{-}$)
and creation operators $b^{\dag}_{\vk}$ for antiparticles. Bogolubov
transformation has a slightly different form in this case,

\bea
a_{\vk}(t)=u\left[ d_{\vk}+a_{\vk}(0)\right]
         +v\left[ \tilde d^{*}_{-{\vk}}+b^{\dag}_{-{\vk}}(0)\right]\nonumber \\
b_{\vk}(t)=u\left[\tilde d_{\vk}+b_{\vk}(0)\right]
          +v\left[d^{*}_{-{\vk}}+a^{\dag}_{-{\vk}}(0)\right]
\label{eq:24}
\eea
containing function $\tilde d_{\vk}$ which can be found from $d_{\vk}$
by the substitution \\ $j({\bf x},t){\to}j^{*}({\bf x},t)$.

As a result single-particle distributions of charged particles have
the same form as those of neutral ones and two-particle distributions
take the form
\be
\langle a^{\dag}_{1}a^{\dag}_{2}a_{1}a_{2}\rangle =
\langle a^{\dag}_{1}a_{1}\rangle \langle a^{\dag}_{2}a_{2} \rangle +
\langle a^{\dag}_{1}a_{2}\rangle \langle a^{\dag}_{2}a_{1} \rangle
\label{eq:25}
\ee
for identical (like-sign) pions and
\be
\langle a^{\dag}_{1}b^{\dag}_{2}a_{1}b_{2} \rangle =
\langle a^{\dag}_{1}a_{1}\rangle \langle b^{\dag}_{2}b_{2} \rangle +
\langle a^{\dag}_{1}b^{\dag}_{2}\rangle \langle a_{1}b_{2} \rangle
\label{eq:26}
\ee
for particle-antiparticle $({\pi}^{+}{\pi}^{-})$ pair with
\bea
\langle a^{\dag}_{1}a_{2} \rangle =
\cosh(r_{1}+r_{2}-\bar r_{1}-\bar r_{2})
\frac {1}{2\sqrt{\bar E_{1}\bar E_{2}}}
\langle j^{*}(\bar E_{1},{\vko})j(\bar E_{2},{\vkt}) \rangle \nonumber \\
+\sinh r_{1}\sinh r_{2} F({\vko}-{\vkt})
\label{eq:27}
\eea

\bea
\langle a_{1}b_{2} \rangle =
\sinh(r_{1}+r_{2}-\bar r_{1}-\bar r_{2})
\frac {1}{2\sqrt{\bar E_{1}\bar E_{2}}}
\langle j^{*}(\bar E_{1},{\vko})j(\bar E_{2},-{\vkt})\rangle \nonumber \\
+\frac{1}{2}\sinh(r_{1}+r_{2})F({\vko}+{\vkt}){}{}{}{}{}{}{}{}{}{}
\label{eq:28}
\eea
(we suggest again that the fields are absent at the initial moment $t=0$).

So like-sign pions have enhanced HBT correlations with additional ground
state contribution (maximal at ${\vko}={\vkt}$ ) and unlike-sign pions
acquire back-to-back correlations ( maximal at ${\vko}=-{\vkt}$) due to
effect of source evolution. The terms of the form $<a_{1}a_{2}>$ are
suppressed for charged pions and they were omitted in the above equations.
The resulting HBT and PAC contributions to the correlation functions
of the charged pairs have the same form as for neutral pions in the previous
section.

The physical interpretation of the above results is clear from the
structure of the Bogolubov transformation. Quasiparticles in the source
partly consist of free pairs (particle having momentum ${\vk}$ and
antiparticle having momentum $-{\vk}$~). The pairs have opposite momenta
of their constituents not to contribute to the total momentum
of the quasiparticle (but influence its energy). The same is valid for
the ground state of the hadronic medium (let us remind that we take
into account time variation of the homogeneous source, say its selfsimilar
expansion). The pairs release when the system decays into free particles.
The HBT effect amplification is also dependent on these pairs.

The origin of doubled correlations between neutral pions is also clear.
Neutral pions are identical particles and simultaneously they are
antiparticles to themselves. So they show both types of correlations.

\section{Conclusions}
The evolution of the source in the course of the particle production
process leads to enhancement of the production rates and to
corresponding amplification of HBT effect characteristic for identical
particles. Moreover the evolution leads to appearance of back-to-back
particle-antiparticle correlations (PAC) which are not suppressed in general.
The width of the corresponding peak (at ${\vko}=-{\vkt}$ ) is close
to that of the HBT peak (at ${\vko}={\vkt}$) and its height is given mainly
by the evolution parameter $r$. In turn the evolution parameter
(which also determines the enhancement of single-particle distributions
and HBT effect) depends on the spectrum of the quasiparticles in the source
and on time duration of the production process, being larger for
soft particles and for fast processes. This parameter can be determined
as soon as PAC will be found in experiment.

Let us note in conclusion that the above effects are of rather general nature.
Field evolution in the expanding Universe~\cite{BD}, Casimir effect
in the volume with moving boundaries and particle production in strong electric
field~\cite{GMM} are described in a similar way using Bogolubov transformation.
These useful analogies are welcomed.

\section*{Acknowledgments}
I would like to thank the participants of the 8th International Workshop
on Multiparticle Production (Matrahaza, Hungary, June 14-21, 1998) for
discussion of the work. I am also indebted to Prof. M.Biyajima who drew
my attention to related paper by H.Hiro-Oka and H.Minakata.
 The work has been supported by the Russian Fund for
Fundamental Research (grant 96-02-16210a).


\end{document}